\documentstyle[12pt]{article}
\begin{document}

\begin{center}
\bigskip
\vspace{.2in}
{\Large \bf A Sober Assessment of Cosmology \\at the New Millennium}

\bigskip
\vspace{.2in}
Michael S. Turner$^{1,2}$\\
\vspace{.2in}
{\it $^1$Departments of Astronomy \& Astrophysics and of Physics\\
Enrico Fermi Institute, The University of Chicago, Chicago, IL~~60637-1433}\\
\vspace{0.1in}
{\it $^2$NASA/Fermilab Astrophysics Center\\
Fermi National Accelerator Laboratory, Batavia, IL~~60510-0500}\\
\end{center}
\pagestyle{plain}
\setcounter{page}{1}

\section{Prologue}

Cosmology is the dot com of the sciences.  Boom or bust.
It is about nothing less than the origin and evolution of the
Universe, the all of everything.  It is the boldest of
enterprises and not for the fainthearted.  Cosmologists are
the flyboys of astrophysics, and they often live
up to all that image conjures up.

It is hardly surprising then that words
written about cosmology are rarely balanced.  They
verve to one extreme or the other -- from, What we can possibly
know about the Universe is a set of measure zero (true); to,
We are on the verge of knowing it all (no comment).  Once or twice,
I have probably been
guilty of irrational exuberance when it comes
to cosmology (see e.g., Turner 1999). (However, I will defend to my last breath
the phrase I coined six years ago, Precision Cosmology.)
Even the most serious scientists have been guilty
of injecting philosophical arguments when discussing cosmology
(cf., the papers of Bondi and Gold (1948) and of Hoyle (1948) on the
aesthetic beauty of the steady state cosmology).

Before I begin my sober assessment of cosmology, let me discuss the
mission statement for the enterprise.
As one might expect, this is where the divergence of views
often begins.  By my definition, cosmology is
the scientific quest to understand the most salient features
of the Universe in which we live.  Since most
of the history of the Universe is uninteresting
thermal equilibrium, the interesting moments, those rare
departures from thermal boringness which lead to
notable features today, are a manageable set.

Others take a much different view.  As an extreme example consider
Disney (2000), who defines cosmology as the quest
to understand the entire space-time history of the Universe
(in response to Disney, see Peebles 2000).
Not surprisingly, he concludes that the achievements of
cosmologists have been minimal and that cosmology may not be a science at all.
While more than two decades ago the relativist Ellis (1975) educated
us all on the impossibility of Disney's goal  --
we are absolutely limited in our knowledge
of the Universe by our past light cone -- that has not prevented
significant progress toward understanding how the basic features of
our portion of the Universe came about as well as their implications
for the Universe as a whole.

My definition of the mission of cosmology is flexible.
It allows the focus to change with time as the science matures.
Today we ask about the nature of the dark matter that
holds the Universe together and the dark energy that
is causing the expansion to speed up.  We are trying to to explain
how all structure we see today, from individual galaxies to
the great walls of galaxies, arose from quantum fluctuations
in the early Universe.
Forty years Sandage described cosmology as the quest for two numbers:
the expansion rate $H_0$ and the deceleration parameter
$q_0$.  Each characterizes the cosmology of its time well.
The evolution of the forefront issues and the
deepening of the questions being asked speaks to
the vitality of cosmology.

I note too that my mission statement for cosmology parallels that
in less controversy-inspiring sciences, e.g.,
astrophysics.  Few astrophysicists would take as the mission
statement for astrophysics the understanding from top to bottom
of each and every object in the Universe.  The aims of astrophysics
are the understanding of classes of objects, persistent themes, 
and general principles.

\section{Cosmology at the Millennium}

The progress made in our understanding of the Universe during the
20th century is nothing short of stunning (see e.g., Turner and
Tyson 1999).  One hundred years
ago the issue of whether or not the laws of physics
applied to the Universe was still being debated.  Today we have
ample evidence for their universality, from the spectra of the
most distant quasars to the nuclear reactions that synthesized
the light elements in the early Universe,
not to mention the success of the big-bang cosmology itself.
At the end of the nineteenth century astronomers
had counted 7 planets, one galaxy and a few million stars.
The planet count now exceeds 40 -- only 9 orbit
our star -- and the Hubble Space Telescope's (HST)
Hubble Deep Field has revealed millions of
galaxies per square degree (with 100 million or so stars within
each galaxy).  The nineteenth century Universe was static and the fires
that fueled stars were a mystery.  Today
we can trace our evolving Universe to a big-bang beginning
14 billion years ago, and we understand the nuclear furnaces
that power stars so well, that in the case of our sun, the central
temperature was correctly predicted to a few percent
(as verified by helioseismology and the detection of solar neutrinos).

Technology has played no small role.  At the beginning of the last century,
my university's 40-inch refractor on the shore of Lake Geneva,
WI was the largest telescope in the world.  The best
seeing was an arcsecond and photographic plates, which
had only recently been introduced, collected less than one
percent of the incident light.  Today, the Keck telescopes with
their 10-meter mirrors sit high atop Mauna Kea, equipped with
CCD detectors that collect nearly all the incident light
and adaptive optics that aspire for HST-class
seeing of 0.1 arcsecond.  Our eyes on the Universe
now span the range from km-length radio waves to gamma-rays of energies $10^{10}$
times that of visible light.  And we have new eyes that
include neutrinos, cosmic rays and soon we hope, gravity waves.

What do we know today with confidence about the Universe?
Here I am referring to facts that have and will stand
the test of time (as in all science, their interpretation
may change as our understanding deepens).  I count four
basic features, all neatly embodied in the hot big-bang model
of general relativity.  Quite properly referred to
as the standard cosmology,\footnote{So named by Steven
Weinberg (1972).  The moniker, standard cosmology,
predates the particle physics equivalent, standard model, by almost
a decade.} the hot big-bang model
is 20th century cosmology's crowning achievement.

{\em We live in an evolving Universe that emerged from a big-bang beginning
some 14 billion years ago.}  The evidence for this is numerous
and varied, including the universal recession of galaxies
(some 300,000 and counting), studies of the birth of galaxies in
the high-redshift Universe, the existence of the cosmic
microwave background (CMB), and the consistency of the ages of
the oldest stars within our galaxy with the time back to
the bang.  Regarding the final point, a variety of techniques
for determining the Hubble constant are now all consistent and pin it down
to 10\% precision, $H_0=70\pm 7\,{\rm km\,sec^{-1}\,Mpc^{-1}}$
(see e.g., Mould et al 2000).   One of Sandage's goals has been accomplished.

{\em During its first million years the Universe was a hot, dense
plasma.}  The evidence for this comes from the
existence of the $2.725\pm 0.001\,$K cosmic microwave background
with its very nearly perfect black-body spectrum (Mather et al, 1999).
(This remarkable measurement made by the FIRAS instrument on NASA's Cosmic
Background Explorer (COBE) satellite is not only
an example of precision cosmology, but also of high-precision science.)
All efforts to produce such a perfect black body by a mechanism
other than an early hot, dense plasma phase have fallen short.

{\em The abundance of cosmic structure seen in the Universe today,
from individual galaxies to superclusters and the great walls
of galaxies, emerged from the tiny (0.001\%) primeval inhomogeneity
in the distribution of matter.}  The evidence comes
from the tens of microKelvin fluctuations in the
temperature of the CMB across the sky.  These ripples in the
microwave sky first revealed by the DMR
instrument on COBE (Smoot et al, 1992) have now been measured
by more than twenty experiments, on angular scales from less than
$0.1\,$degree to 100\,degrees.  They map the slightly lumpy distribution
of matter at the time of last scattering (400,000
years after the big bang).\footnote{I note that the precision
in measuring the lumpiness of the matter distribution,
better than 0.001\%, exceeds the official margin between Bush and
Gore in Florida by a factor of 5.  Precision cosmology!}
Taking into account the gravitational
amplification of the inhomogeneity in the 14 billion years since,
the level of lumpiness in the
matter is consistent with the structure that exists in the Universe today.

{\em The Universe at an age of seconds, when thermal
energies were in the keV to MeV range, was a nuclear furnace
operating out of thermal equilibrium.}
The evidence comes in the abundance pattern of H (about 76\% by
mass), D (a few parts in $10^{-5}$ relative
to H), $^3$He (similar to D),
$^4$He (about 24\% by mass), and $^7$Li (a few parts in $10^{-10}$
relative to H) seen in the most pristine samples
of the cosmos.  This is the earliest, and perhaps most impressive,
milepost in the known history of the Universe.

From the nuclear oven at 1 second, it is a short extrapolation
back to $10^{-5}\,$sec and
thermal energies of around $100\,$MeV when the
favored state of ordinary matter was a
quark/gluon plasma.
There is no hard evidence for the early quark-soup phase,
in part, because the transition from quarks and gluons
to neutrons and protons occurred in thermal equilibrium
(or close to it).  However, strong indirect
support comes from detailed numerical simulations of the quark/hadron
transition (lattice gauge theory) and Quantum Chromodynamics Dynamics,
the theory of the strong color force that binds quarks
into neutrons, protons and the other hadrons.  More direct evidence may 
soon come from the laboratory creation
of quark/gluon plasma at Relativistic Heavy Ion Collider (RHIC)
at Brookhaven National Laboratory.

The uniformity of the CMB across the sky ($\delta T/T < 10^{-4}$)
and the homogeneity in the large seen in the biggest
galaxy redshift surveys
(LCRS with 60,000 redshifts, SDSS with 100,000 redshifts,
and the 2dF survey with 200,000 redshifts) testify to
the isotropy and homogeneity of the observed Universe on the
largest scales.  Theory and Occam's razor would argue that a
much larger portion of the Universe than our past light cone
is similarly smooth and had a similar history.
But Ellis reminds us that this is not a cosmological fact.
The smoothness issue illustrates well the tension between the absolute
mathematical limitations of cosmology and reasonable,
though not rigorous, grander inferences that we would like to make.

Twenty years ago most cosmologists interpreted the homogeneity and
isotropy of the observable universe as evidence for
the smoothness of the Universe as a whole, based upon
the Cosmological Principle (principle of mediocrity), which states that
we do not live in a special place.  Such an interpretation fits
nicely within the homogeneous and isotropic Friedmann-Robertson-Walker
(FRW) cosmological solution.  Collins and Hawking (1973)
among others raised the issue of the specialness of the
FRW solution (even the class of slightly lumpy
FRW solutions occupies only a set
of measure zero in the space of initial data).
Misner and others launched a program to see if
the isotropy and homogeneity could have
arisen in more general solutions through the smoothing
effects of particle interactions.  However, the
smallness of the particle horizon within FRW-like solutions at
early times precludes this kind of smoothing.
A completely different approach -- Guth's inflationary Universe --
did lead to a possible resolution and a very different big picture.

According to the inflationary view, our smooth observable
Universe sits well within a much, much larger smooth region,
that of our inflationary bubble.
The Universe as a whole consists of an infinity
of such bubbles and thus has a multiverse structure.
This view, like its predecessor, is perfectly consistent
with all observations.  Not only is it radically different,
it is at odds with the Cosmological Principle.  The individual
bubble universes can differ greatly, e.g., in the number of
spatial dimensions or the kinds of particles that are stable,
and when viewed on the grandest of scales the Universe is not
described by the FRW cosmology.

The standard of proof in cosmology will never be that
of laboratory science.  There will be ideas that
may never be directly tested; for example, the multiverse
structure of space time predicted by inflation.
Nonetheless, the standard of proof can be impressive.
Take for example, big-bang nucleosynthesis.
It began with Gamow's grand idea of the early Universe as a nuclear
oven that cooked all the elements in the periodic table.  Its
very first prediction, the existence of the cosmic microwave
background was confirmed.\footnote{The story of the very
loose link between the prediction of the CMB by Gamow and
his collaborators and its accidental discovery by Penzias
and Wilson is well known (see e.g., Weinberg, 1977).}  Theorists including
Hayashi, Fermi, Turkevich, Alpher, Herman, Follin, Hoyle, Peebles,
Fowler, Wagoner, and my late colleague
David Schramm all filled in crucial details.  All of the
important nuclear reactions have been studied and measured
in the lab at the relevant energies.

Today, the theoretical uncertainties
in the predicted abundances are at at the 0.15\% level for
$^4$He, 3\% level for D and $^3$He and 15\% for $^7$Li.
The observational data is approaching this level of precision
too.  The primeval deuterium abundance has been measured
in high redshift ($z\sim 2 - 4$) hydrogen clouds to 
10\%; the primeval helium abundance is known
to about 5\% in metal-poor galaxies; and the Lithium abundance
is determined to around 30\% in old pop II halo stars.
While significant issues remain and more precision is possible,
the state of affairs is quite good (see e.g., Schramm and Turner,
1998 or Olive et al, 2000).  Big-bang nucleosynthesis shows just how
well cosmological facts can be established and
helps to make the case for precision cosmology.

\section{Looking Ahead}

The cosmological clock ticks logarithmically.  The standard
cosmology has filled in the the most recent 20 ticks.
The earliest 40 ticks remain to be studied,
and there are indications that some of the most
fundamental features of the Universe today trace their
origins to this epoch.
As successful as it is, the hot big-bang model
leaves unanswered a number of important questions -- origin of the smoothness,
tiny primeval matter lumpiness and ordinary matter itself, 
the nature of the dark matter and dark energy,
and what went bang -- and raises the expectations for what
can be learned from the early Universe.

The realization in the 1970s that the fundamental
particles are point-like quarks and leptons and not finite-sized
hadrons like neutrons and protons (which would have
overlapped one another at $10^{-5}\,$sec) opened the door to
sensible speculation about the universe all the way back
to the Planck time ($10^{-43}\,$sec).  During the 1980s
many bold and intriguing ideas about the first 40 ticks based upon
speculative ideas about unification of the elementary particles
and forces were put forth and provided intellectual fuel to
help power the next cosmological boom (see e.g., Kolb and Turner, 1990)

In the 1990s the next generation of forward-looking Gamow-like ideas
ready for testing were culled from the previous decade of ``intense
theoretical speculation'' (referred to by some as the go-go junk
bond days of early-Universe cosmology).  They include:
elementary particles as dark matter, cosmological inflation,
baryogenesis, cosmological phase transitions, and cosmological leftovers
such as cosmic string and monopoles.  Of all of these,
inflation and cold dark matter are the most expansive
and compelling.  In a moment of exuberance,
I would venture to say that these two ideas, as
much as technological advances, have helped to spur
the impressive program of experiments and observations
slated for the next 20 years.

Over the past few years we have had a glimpse of what the future
holds for cosmology in the next century -- and it looks pretty exciting.
Measurements of fine-scale CMB anisotropy by BOOMERanG, MAXIMA and other experiments
have made a strong case for a flat Universe (deBernardis et al 2000;
Hanany et al 2000; Knox and Page 2000).
With the evidence presented by the Supernova Cosmology Project 
(Perlmuter et al 1999) and the High-z Supernova Search Team 
(Riess et al 1999) for accelerated expansion,
we now have a tentative accounting of all the stuff in the Universe:
1/3rd is matter (5\% ordinary matter and 30\% or so exotic dark matter)
and 2/3rds is dark energy (stuff with large, negative
pressure that is leading to accelerated expansion).
These measurements gave a boost to inflation and presented a
new puzzle -- 2/3rds of the Universe is stuff more
mysterious than Zwicky's dark matter.

This accounting speaks to the state of cosmology
today -- significant progress toward answering very basic
questions but important issues still unresolved and
room for more surprises.  ``Conventional cosmology''  today with
its dark matter and dark energy is very much out on a limb.
According to Karl Popper that's what strong theories do!

The case for the existence of dark matter is firmly
rooted in Newtonian physics:  The gravity of
luminous matter is not sufficient to hold structures
in the Universe together.  The argument for
nonbaryonic dark matter is tied to
the baryon density determined from BBN and CMB anisotropy
(see e.g., Burles et al 2001)
which falls a factor of at least 5 short of the needed
amount of dark matter.  While no self-consistent
modification of Newtonian/Einsteinian gravity
can eliminate the need for dark
matter, and the idea of particles from
the early Universe as dark matter is both theoretically
compelling and at least partially correct (neutrinos
have sufficient mass to account for at least as much mass as do
stars), the solution to the dark-matter problem could
involve a modification of gravity.
The acceleration of the expansion
of the Universe is easily accommodated by Einstein's
theory of gravity (in general relativity gravity can be repulsive).
Today we look to dark energy, something akin (or even
identical to) Einstein's cosmological constant
as the explanation (see e.g., Turner 2000).  However, Einstein's
general relativity could be the culprit.

Either way, the solutions to the
dark matter and dark energy problems will deepen
the connections that already exist
between fundamental physics and cosmology.

According to my definition, cosmology is about explaining and
understanding the fundamental features of the Universe.  Inflation
+ cold dark matter is a bold attempt to greatly extend our
understanding.  It holds that
we live in a smooth, spatially flat
bubble universe whose ``big-bang event'' was the enormous
burst of expansion associated with inflation.  The
smoothness of the observable universe arises because
it grew from an extremely tiny region of space, and its
flatness is due to flattening effect of that same
tremendous expansion.  If correct,
the primeval matter lumpiness arose from quantum fluctuations
associated with the inflationary phase, and the matter that
holds all structures together is made primarily of slowly moving
elementary particles (the cold dark matter)
left over from the earliest moments of particle democracy.

Inflation + cold dark matter is bold and testable.  Its
key predictions are the flatness of the universe,
the signature for the quantum origin of lumpiness in the
acoustic peaks in the CMB anisotropy, the pattern
of structure formation predicted by the
cold dark matter scenario, and a stochastic
background of gravity waves.

The first three predictions are already being addressed
by numerous observations, and inflation + cold dark matter
has fared well so far.
Over the next twenty years, the variety and sharpness
of the tests will grow significantly.  Detection of the
``smoking-gun signature''  of inflation -- gravity waves
produced quantum fluctuations in the metric during inflation --
is a key challenge for the next century.

One measure of a subject are its
big questions.  In cosmology today they include:
\begin{itemize}

\item Will the standard hot big bang framework
(including general relativity) hold up to the precision measurements
that will made over the next twenty years?

\item What is the name of the particle that makes up the dark matter
which holds structures in the Universe together?

\item What is the nature of the dark energy that is causing the Universe
to speed up and how long will the accelerated expansion continue?

\item Did the Universe undergo a period of inflation and if so
what is the underlying physical cause?

\item How did the ordinary matter come into existence?

\item What is the explanation for the strange cosmic
mix of baryons, cold dark matter, neutrinos and dark energy?

\item What went bang?

\end{itemize}
All of these questions will be addressed by the dazzling
array of observations and experiments coming in the
next two decades.  Many of them will also be answered.

Where might cosmology stand 20 years from now?  Precision
cosmological tests could put both the standard
cosmology, and its theoretical foundation, general relativity,
on a much firmer footing.  Included in these
are independent 1\% measurements
of the baryon density from light-element abundances and CMB
anisotropy; crosschecks of the Hubble constant from several
different direct determinations as well as an indirect
determination from CMB anisotropy; and detailed comparison
of the structure of the CMB anisotropy with that predicted by theory.

Particle dark matter could become well enough established for
the most skeptical astronomer to believe in!  The neutralinos
holding our own galaxy together could be directly detected
by one of the several laboratory experiments now running
(e.g., the CDMSII experiment) and directly produced
in proton-antiproton collisions at the Fermilab Tevatron during
the upcoming run or later at the LHC.  [While the detection
of the other leading candidate, the axion, could not as
easily be followed up by laboratory detection, its
signature in the conversion of axions to microwave photons
is especially distinctive.]

I do not yet see a suite of measurements that could
convince the most skeptical astronomer that the Universe
really inflated.  (I hope I am just shortsighted.)
To illustrate my worry,
suppose that MAP and Planck measurements of the CMB are
consistent with inflation in every way, and that
the tell-tale signature of inflation, gravity waves with a
nearly scale-invariant spectrum, are detected as well.  These
measurements would reveal much about the underlying physics
of inflation, even allowing the potential energy curve
of the inflaton field to be determined.  What I don't see
in my crystal ball is a laboratory experiment
that would close the circle as nuclear-physics measurements
did for the big-bang nucleosynthesis, as the creation of
quark-gluon plasma at an accelerator could do for the quark-soup
phase of the early Universe, or as the laboratory production
of dark-matter particles could do for particle dark matter.

What about the rest of the 21st century?  While I am bullish
on cosmology, I worry that without a new window, a
new relic, or powerful new theoretical ideas
cosmology could be in for a dry spell 20 or 30 years
hence.  The Universe we strive to understand
remains barely within the reach
of our most powerful ideas and instruments today.
We were ill prepared for the discovery of the
expansion of the Universe in 1929, both theoretically and technologically,
and a slow period in cosmology followed.  The dawn came
in the 1960s with the advent of the 200-inch Hale telescope,
the discovery of quasars and the CMB, and a better
theoretical understanding of the big bang.

Cosmology is a science (yes indeed!) that asks grand questions
and sometimes finds equally grand answers.  No one can
deny that the job description for cosmologist includes
a bit of arrogance.  It is a field of boom or bust that
sometimes cannot resist philosophizing.
While it doesn't take a cosmologist to tell you
that the next twenty years will be very exciting, not
even a cosmologist would dare predict where cosmology
will be at the beginning of the next century.

\end{document}